\documentclass{JHEP3}
\usepackage{amssymb}
\usepackage{graphicx}
\usepackage{epsfig}
\usepackage{axodraw}
\newcommand{\bmat}{\left(\begin{array}}
\newcommand{\emat}{\end{array}\right)}
\newcommand{\be}{\begin{equation}}
\newcommand{\ee}{\end{equation}}
\newcommand{\bea}{\begin{eqnarray}}
\newcommand{\eea}{\end{eqnarray}}
\def\ie{{\it i.e.}}
\def\lsim{\raise0.3ex\hbox{$\;<$\kern-0.75em\raise-1.1ex\hbox{$\sim\;$}}}
\def\gsim{\raise0.3ex\hbox{$\;>$\kern-0.75em\raise-1.1ex\hbox{$\sim\;$}}}

\title{Lepton flavor violation in supersymmetric $B-L$ extension of the
standard model}

\author{Shaaban Khalil\\
Center for Theoretical Physics at the
British University in Egypt, Sherouk City, Cairo 11837, Egypt.\\
Department of Mathematics, Faculty of Science, Ain Shams
University, Cairo 11566, Egypt.}

\abstract{Supersymmetric $B-L$ extension of the Standard Model
(SM) is one of the best candidate for physics beyond the SM that
accounts for TeV scale seesaw mechanism and provides an attractive
solution for the Higgs naturalness problem. We analyze the charged
lepton flavor violation (LFV) in this class of models. We show
that due to the smallness of Dirac neutrino Yukawa coupling, the
decay rates of $l_i \to l_j \gamma$ and $l_i \to 3 l_j$, generated
by the renormalization group evolution of soft SUSY breaking terms
from GUT to seesaw scale, are quite suppressed. Therefore, this
model is free from the stringent LFV constraints usually imposed
on the supersymmetric seesaw model. We also demonstrate that the
right-sneutrino is a long-lived particle and can be pair produced
at the LHC through the $B-L$ gauge boson. Then, they decay into
same-sign dilepton, with a total cross section of order ${\cal
O}(1)$ pb. This signal is one of the striking signatures of
supersymmetric $B-L$ extension of the SM.}


\preprint{}
\begin{document}
\section{Introduction}
If neutrinos were massless, the Lepton Flavor (LF) in the charged
sector of the SM would be conserved. The observed neutrino
oscillations are evidences for neutrino masses, which may entail
that lepton flavor is no longer conserved. Nevertheless, LFV is
almost forbidden in the SM with massive neutrinos. The processes
of charged LFV are suppressed by tiny ratio of neutrino masses to
the W-boson mass. For instance, the branching ratio of decay $\mu
\to e \gamma$ is of order $ 10^{-43} (m_{\nu}/1 eV)^4$, which is
far from the experimental reach.

In fact, there is no fundamental reason that implies the
conservation of LF in the SM. LF is an accidental symmetry at low
energy, and it may be violated beyond the SM. Indeed, several SM
extensions, like grand unified field theory (GUT), technicolor,
and supersymmetry, indicate the possibility of large LFV.
Therefore, a signal of LFV in charged lepton sector would be a
clear hint for physics beyond the SM. The present experimental
limits \cite{Brooks:1999pu} are: %
\bea
 BR(\mu \to e \gamma) & <& 1.2 \times 10^{-11}~,\nonumber\\
 BR(\tau \to \mu \gamma)& <& 6.8 \times 10^{-8}~,\nonumber\\
 BR(\tau \to e \gamma)& <& 1.1 \times 10^{-7}~,\nonumber\\
 BR(\mu \to 3 e)& <& 1.0 \times 10^{-12}.
\eea %
The MEG experiment at PSI \cite{meg} is expected to reach the
limit of $10^{-13}$ for the branching ratio of $\mu \to e \gamma$
and $\mu \to 3 e $ processes. This will be a very serious test for
physics beyond the SM.

Supersymmetry is an attractive candidate for new physics at TeV
scale that provides an elegant solution for the SM gauge hierarchy
problem and stabilize the SM Higgs mass at the electrwoeak scale.
In SUSY models, new particles and new interactions are introduced
that lead to potentially large LFV rates. Therefore, searches for
LFV in charged sector may probe the pattern of SUSY breaking and
constrain its origin \cite{Carvalho:2000xg}. Furthermore, seesaw
mechanism is an interesting solution to the problem of the small
neutrino masses. In what is called type I seesaw mechanism, SM
singlets ( right-handed neutrinos) with mass of order ${\cal
O}(10^{14})$ GeV are introduced. It turns out that the combination
of these two interesting ideas of SUSY and seesaw implies sizable
rates for LFV, even when SUSY breaking terms are assumed to be
completely flavor blind. Consequently, the SUSY spectrum should be
pushed up to few TeV's \cite{Borzumati:1986qx}. In this case,
there will be no hope to prob SUSY particles at LHC. Also, with
very heavy right-handed neutrino, there is no way to test the
seesaw mechanism directly at the LHC. Therefore, TeV scale seesaw
mechanism was well motivated and has been recently considered as
an alternative paradigm \cite{Khalil:2006yi}.

The TeV scale right-handed neutrino can be naturally implemented
in supersymmeric $B-L$ extension of the SM (SUSY $B-L$), which is
based on the gauge group $G_{B-L}\equiv SU(3)_C \times SU(2)_L
\times U(1)_Y \times U(1)_{B-L}$ \cite{Khalil:2007dr}. In this
model, three SM singlet fermions arise quite naturally due to the
$U(1)_{B-L}$ anomaly cancellation conditions. These particles are
accounted for right-handed neutrinos, and hence a natural
explanation for the seesaw mechanism is obtained
\cite{Khalil:2006yi,Abbas:2007ag,Emam:2007dy}. The masses of these
right-handed neutrinos are of order the $B-L$ breaking scale. In
SUSY $B-L$ model, the $B-L$ Higgs potential receives large
radiative corrections that induce spontaneous $B-L$ symmetry
breaking at TeV scale, in analogy to the electroweak symmetry
breaking in MSSM \cite{Khalil:2007dr}. In this case, to fulfill
the experimental measurements for the light-neutrino masses, with
TeV scale right-handed neutrinos, the Dirac neutrino masses should
be order ${\cal O}(10^{-4})$ GeV, \ie, they have to be as light as
the electron.

In this paper we analyze the LFV in SUSY $B-L$ model. We show that
due to the smallness of Dirac neutrino Yukawa couplings, the decay
rate of of $l_i \to l_j \gamma$ and $l_i \to 3 l_j$ are quite
suppressed. Hence, the predictions of SUSY $B-L$ for the branching
ratio of these processes remain identical to the MSSM ones. Also,
we study the pair production of right-sneutrinos at the LHC and
show that they are long-lived particles. The decay of these
right-sneutrinos leads to a very interesting signal of the
same-sign dilepton with possible different lepton flavors. We
demonstrate that the cross section of this event is of order
${\cal O}(1)$ pb. Therefore, it is quite accessible at the LHC and
can be considered as indisputable evidence for SUSY $B-L$ model.

The paper is organized as follows. In section 2 we present the
main features of the SUSY $B-L$ extension of the SM. In
particular, we analyze the spontaneous $B-L$ breaking at TeV scale
by large radiative corrections to the $B-L$ Higgs potential.
Section 3 is devoted for the LFV in SUSY $B-L$. We start with the
conventional $l_i \to l_j$ transitions, then we study the
same-sign dilepton event which is a clean signal for large
right-sneutrino mixing. Finally we give our conclusions in section
4.


\section{Supersymmetric $B-L$ extension of the SM}

In this section we analyze the minimal supersymmetric version of
the $B-L$ extension of the SM based on the gauge group
$G_{B-L}=SU(3)_C \times SU(2)_L \times U(1)_Y \times U(1)_{B-L}$.
This SUSY $B-L$ is a natural extension of the MSSM with three
right-handed neutrinos to account for measurements of light
neutrino masses. The particle content of the SUSY $B-L$ is the
same content as the MSSM with the following extra particles: three
chiral right-handed superfields ($N_i$), vector superfield
necessary to gauge the $U(1)_{B-L}$ ($Z_{B-L}$), and two chiral
SM-singlet Higgs superfields ($\chi_1$, $\chi_2$ with $B-L$
charges $Y_{B-L}=-2$ and $Y_{B-L}=+2$ respectively). As in MSSM,
the introduction of a second Higgs singlet ($\chi_2$) is necessary
in order to cancel the $U_{B-L}$ anomalies produced by the
fermionic member of the first Higgs ($\chi_1$) superfield.

 The interactions between Higgs and matter superfields are
described by the superpotential %
\begin{eqnarray}%
W_{B-L} = W_{MSSM} +(Y_\nu)_{ij} L_i H_2 N_j^c + (Y_N)_{ij}
N^c_i N^c_j \chi_1 + \mu H_1 H_2 + \mu^\prime \chi_1 \chi_2.%
\label{superpot}%
\end{eqnarray}
Note that $Y_{B-L}$ for leptons and Higgs are given by
\cite{Khalil:2006yi}: %
\bea %
Y_{B-L}(L)= Y_{B-L}(E)=Y_{B-L}(N)=-1,~~~~~~~~   Y_{B-L}(H_1)= Y_{B-L}(H_2)=0.%
\eea %
It also remarkable that due to the $B-L$ gauge symmetry, the
$R$-parity violating terms are now forbidden. These terms violate
baryon and lepton number explicitly and lead to proton decay at
unacceptable rates. On the other hand, the relevant soft SUSY
breaking terms, assuming certain universality of soft SUSY
breaking terms at GUT scale are in general  given by
\begin{eqnarray}
- {\cal L}^{B-L}_{soft} &=&-{\cal L}^{MSSM}_{soft} +
{\widetilde{m}}_{Nij}^{2}{\widetilde{N}}_{i}^{c*
}{\widetilde{N}}_{j}^{c} + m^2_{\chi_1} \vert{\chi_1}\vert^2 +
 m^2_{\chi_2}\vert{\chi_2}\vert^2 \nonumber\\
 &+& \left[ Y_{\nu ij}^{A}{\widetilde{L}}_{i}
{\widetilde{N}^c}_{j}H_{u} + Y_{N ij}^{A}{\widetilde{N}}_i^{c}
{\widetilde{N}}_j^{c}\chi_{1} + B \mu^\prime \chi_1 \chi_2 +
\frac{1}{2} M_{B-L}{\widetilde{Z}_{B-L}}{\widetilde{Z}_{B-L}}+ h.c
\right] ,~~ %
\label{Lsoft}%
\end{eqnarray}%
where $(Y_N^A)_{ij}\equiv (Y_N A_N)_{ij}$ is the trilinear
associated with Majorana neutrino Yukawa coupling. We now show how
the $B-L$ breaking scale can be related to the scale of SUSY
breaking, as emphasized in Ref.\cite{Khalil:2007dr}. The scalar
potential for the Higgs fields $H_{1,2}$ and $\chi_{1,2}$ is given
by
\begin{eqnarray}
V(H_1,H_2,\chi_1,\chi_2) & = & \frac{1}{2}
g^2(H_1^*\frac{\tau^a}{2} H_1 + H_2\frac{\tau^a} {2}H_2)^2 +
\frac{1}{8} g^{\prime^2}(\vert H_2\vert^2 - \vert H_1\vert^2)^2
        \nonumber\\
            & + & \frac{1}{2} g^{\prime\prime^2}(\vert \chi_2\vert^2 - \vert \chi_1\vert^2)^2 + m_1^2 \vert H_1\vert^2 +
            m_2^2\vert H_2\vert^2 - m_3^2(H_1 H_2 + h.c) \nonumber\\
            & + & \mu^2_1 \vert \chi_1\vert^2 +
            \mu^2_2\vert \chi_2\vert^2 - \mu^2_3(\chi_1 \chi_2 + h.c)  ,
\label{Vtotal}
\end{eqnarray}
where
\begin{eqnarray}
m^2_i &=& m_0^2 + \mu^2 , \hspace{0.5cm} i=1,2 \hspace{2cm} m_3^2
=- B \mu~, \label{m12}\\
\mu^2_i &=& m_0^2 + \mu^{\prime^2} , \hspace{0.5cm} i=1,2
\hspace{2cm} \mu^2_3 = - B \mu^\prime~.
\label{mp12}
\end{eqnarray}
As can be seen from Eq.(\ref{Vtotal}), the potential
$V(H_1,H_2,\chi_1,\chi_2)$ is factorizable. It can be written as
$V(H_1,H_2) + V(\chi_1,\chi_2)$ where $V(H_1,H_2)$ is the
usual MSSM scalar potential which leads to the radiative
electroweak symmetry breaking. As is known, due to the running
from GUT to weak scale with large top Yukawa coupling, $m_2^2$
receives negative contributions that radiatively breaks the
electroweak symmetry. Therefore, we will focus here on the new
potential $V(\chi_1,\chi_2)$ to analyze the possibility of
breaking $B-L$ at TeV scale, through the soft SUSY breaking terms.
This potential is given by %
\be %
V(\chi_1,\chi_2)= \frac{1}{2} g^{\prime\prime^2}(\vert
\chi_2\vert^2 - \vert \chi_1\vert^2)^2 + \mu^2_1 \vert
\chi_1\vert^2 + \mu^2_2\vert \chi_2\vert^2 - \mu^2_3(\chi_1 \chi_2
+
h.c).%
\ee %
It should be noted that $V(\chi_1,\chi_2)$ is quite similar to the
MSSM Higgs potential which spontaneously breaks the electroweak
symmetry. Therefore it is expected that the $B-L$ symmetry
breaking approach is going to be the same as the well known
procedure of elctroweak symmetry breaking in MSSM. The
minimization of
$V(\chi_1,\chi_2)$ leads to the following condition:%
\be %
v^{\prime^2} = (v^{\prime^2}_1 + v^{\prime^2}_2) =
\frac{(\mu^2_1-\mu^2_2) - (\mu^2_1+\mu^2_2)\cos2\theta}{2
g^{\prime\prime^2}
\cos2\theta},%
\label{muprime}
\ee %
where $\langle \chi_1 \rangle = v_1$ and $\langle \chi_2 \rangle =
v_2$. The angle $\theta$ is defined as $\tan \theta =v_1/v_2$. The
minimization conditions also leads to
\be %
\sin 2 \theta = \frac{2
\mu^2_3}{\mu^2_1+\mu^2_2}.%
\ee %

After $B-L$ breaking, the $Z_{B-L}$ gauge boson acquires a mass
\cite{Khalil:2006yi}: $M^2_{Z_{B-L}}=4 g^{\prime\prime^2}
v^{\prime^2}$. The high energy experimental searches for an extra neutral
gauge boson impose lower bounds on this mass. The stringent
constraint on $U(1)_{B-L}$ obtained from LEP ll result, which
implies \cite{Carena:2004xs} %
\be
\frac{M_{Z_{B-L}}}{g^{\prime\prime}}>6TeV.\label{z-b-l-constrain} %
\ee
The discovery potential for $Z_{B-L}$ at the LHC has been analyzed
through its decay into an electron--positron pair
\cite{Emam:2008zz} and into 3 leptons \cite{Basso:2009hf}. It was
shown that ${\cal O}(1)$ TeV $Z_{B -L}$ can be easily probed at
the LHC with an integrated luminosity of order $\sim 100$
fb$^{-1}$.

For a given $M_{Z_{B-L}}$, the minimization
condition (\ref{muprime}) can be used to determine the
supersymmetric
parameter $\mu^{\prime^2}$, up to a sign. One finds %
\be %
\mu^{\prime^2}= \frac{m_{\chi_2}^2 - m_{\chi_1}^2 \tan^2 \theta }{\tan^2\theta-1} - \frac{1}{4}M_{Z_{B-L}}^2 . %
\ee%
In order to ensure that the potential $V(\chi_1,\chi_2)$ is
bounded from below, one must require
\begin{equation}
\mu_1^2 + \mu_2^2 > 2 \vert \mu_3^2\vert ~. %
\label{stab}
\end{equation}
This is the stability condition for the potential. Also, to avoid
that $\langle \chi_1 \rangle = \langle \chi_2 \rangle  = 0$ be a
local minimum we have to require
\begin{equation}
\mu_1^2 \mu_2^2 < \mu_3^4~ . \label{minimiz}
\end{equation}
It is clear that with positive values of $\mu_1^2$ and $\mu_2^2$,
given in Eq.(\ref{mp12}), one can not simultaneously fulfill the
above conditions. However, as pointed out in Ref.
\cite{Khalil:2007dr}, the renormalization group evolutions of the
scalar masses $m_{\chi_1}^2$ and $m_{\chi_2}^2$ of Higgs singlets
$\chi_1$ and $\chi_2$ are different. Therefore, at TeV scale the
mass $m^2_{\chi_1}$ becomes negative, whereas $m^2_{\chi_2}$
remains positive. In this case, both of the electroweak, $B-L$ and
SUSY breakings are linked at scale of ${\cal O}$(TeV).

In this regards, the observed light-neutrino masses can be
obtained if the neutrino Yukawa couplings, $Y_{\nu}$, are of order
${\cal O}(10^{-6})$ \cite{Khalil:2006yi,Abbas:2007ag}, which are
close to the order of magnitude of the electron Yukawa coupling.
The LHC discovery for TeV right-handed neutrino in $B-L$ extension
of the SM has been studied in Ref.\cite{Huitu:2008gf}. It was
shown that the production rate of the right-handed neutrinos is
quite large over a significant range of parameter space. Searching
for the right-handed neutrinos is accessible via four lepton
channel, which is a very clean signal at LHC, with negligibly
small SM background.

With TeV scale right-sneutino, the low-energy sneutrino mass
matrix is given by $12\times 12$ hermitian matrix
\cite{Kajiyama:2009ae}. However the mixing between left- and
right- sneutrinos is quite suppressed since it is proportional to
Yukawa coupling $Y_\nu \lsim {\cal O}(10^{-6})$. A large mixing
between the right-sneutrinos and anti- right-sneutrinos is quite
plausible, since it is given in terms the Yukawa $Y_N \sim {\cal
O}(1)$. Therefore, one can focus on the right-sneutrino sector and
study the possible oscillation between sneutrino and
anti-sneutrino. The right-sneutrino mass matrix in the $(\tilde N^c, \tilde
N^{c*})$ basis can be written as%
\be %
{\cal M}^2 \simeq \left(\begin{array}{cc}
                \tilde{m}_N^2 + M_N^2 & - v'_1~ (Y_N^A)^*+ v'_2 Y_N \mu'\\
                - v'_1~ (Y_N^A) + v'_2~ Y_N \mu'^* & \tilde{m}_N^2 + M_N^2)
                \end{array}\right).
\ee %
As can be seen from the above expression, the off-diagonal
elements could be of the same order as the diagonal ones.
Therefore, a large mixing can be obtained. In this case, the
$(6 \times 6)$ right-sneutrino mass matrix is diagonalized by unitary matrix $X_{\tilde{\nu}}$: %
\be %
X_{\tilde{\nu}} {\cal M}^2 X_{\tilde{\nu}}^\dag = {\cal
M}^2_{diag}. %
\ee %
Hence, %
\be %
\tilde{\nu}_{R_i} = (X_{\tilde{\nu}})_{ij}
\tilde{N}_j~, ~~~~~ i,j=1,2...,6.%
\ee %

Finally, we consider the neutral gaugino and Higgsino sector which is
going to be modified by the new $B-L$ gaugino and the fermionic
partners of the singlet scalar $\chi_{1,2}$. In the weak
interaction basis defined by $\psi^{0^T}=\left(\tilde{B}^0,
\tilde{W}_3^0, \tilde{H}_1^0,
\tilde{H}_2^0,\tilde{\chi}_1^0, \tilde{\chi}_2^0,\tilde{Z}^0_{B-L}\right)$, the
neutral fermion mass matrix is given by the following $7\times 7$
matrix \cite{Khalil:2008ps}: %
\bea &&{\cal M}_n = \left(\begin{array}{cc}
{\cal M}_4 & {\cal O}\\
 {\cal O} &  {\cal M}_3\\
\end{array}\right),
\eea%
where the ${\cal M}_4$ is the MSSM-type neutralino mass matrix and
${\cal M}_{3}$ is the additional neutralino mass matrix with
$3\times 3$:%
\bea %
{\cal M}_3 &=& \left(\begin{array}{ccc}
0 & -\mu' & -2g^{\prime\prime}v'\sin\theta \\
-\mu' & 0 & 2g^{\prime\prime}v'\cos\theta  \\
-2g^{\prime\prime}v'\sin\theta & +2g^{\prime\prime}v'\cos\theta & M_{1/2}\\
\end{array}\right).
\label{mass-matrix.1} \eea
In case of real mass matrix, one diagonalizes the matrix ${\cal
M}_{n}$ with a symmetric mixing matrix $V$ such as
\be V{\cal M}_n
V^{T}=diag.(m_{\chi^0_k}),~~k=1,..,7.\label{general} \ee In this
aspect, the lightest neutralino (LSP) has the following
decomposition \be \chi^0_1=V_{11}{\widetilde B}+V_{12}{\widetilde
W}^3+V_{13}{\widetilde H}^0_d+V_{14}{\widetilde
H}^0_u+V_{15}{\widetilde \chi_1}+V_{16}{\widetilde
\chi_2}+V_{17}{\widetilde Z}_{B-L}. \ee The LSP is called pure
$\widetilde Z_{B-L}$ if $V_{17}\sim1$ and $V_{1i}\sim0$,
$i=1,..,6$ and pure $\widetilde\chi_{1(2)}$ if $V_{15(6)}\sim1$
and all the other coefficients are close to zero
\cite{Khalil:2008ps}.
It is worth noting that the MSSM chargino mass matrix remains
intact in this type of models, since there is no any new charged
fermion have been introduced.

\section{LFV in SUSY $B-L$ model}
In MSSM, the SUSY contributions to the decay channels of $l_i \to
l_j \gamma$ are dominated by one loop diagrams with
neutralino-slepton and chargino-sneutrino exchanges. It turns out that
the experimental limit on $\mu \to e \gamma$ induces stringent
constraints on the transitions between first and second
generations. Applying the $\mu \to e \gamma$ constraints on the
neutralino contribution leads to the following upper bounds of the slepton mass insertions : %
\be%
(\delta^l_{LR})_{12} \lsim 10^{-6}, ~~~  ~~~  ~~~
(\delta^l_{LL})_{12} \lsim 10^{-3}. %
\label{bounds}
\ee%
Note that due to the $SU(2)_L$ gauge invariance, one gets the
following relation between slepton and sneutrino mass insertions:
$(\delta^\nu_{LL})_{ij} \simeq (\delta^l_{LL})_{ij}$.
For $(\delta^\nu_{LL})_{12} \simeq 10^{-3}$, the chargino contribution
to $\mu \to e \gamma$ is automatically below the current
experimental limit. These bounds generally impose very stringent constraints
on the soft SUSY breaking terms, known as SUSY flavor problem.

It is also worth mentioning that $\mu
\to 3 e$ and $\mu \to e$ conversion in nuclui, \ie, $\mu + N \to e
+ N$ are considered as another source of probing possible SUSY
effects. The relation between these two processes and $\mu \to e
\gamma$ is, in general, model independent. However, in SUSY framework, where
these processes are generated by the photon penguin, $Z$-penguin
and box diagrams, one usually finds $ BR(\mu \to 3 e) \sim BR(\mu
\to e) \sim {\cal O} (10^{-3}) \times BR(\mu \to e \gamma)$. In this
respect, it seems the present limit on $\mu \to 3 e$ and $\mu \to
e$ conversion are less sensitive than the current bound on $\mu
\to e \gamma$. However, future experiments would reach the limit
of $10^{-17}$ for the branching ratio of $\mu \to e$ conversion
and $10^{-16}$ for $BR(\mu \to 3 e)$, while $BR(\mu \to e \gamma)$
may approach $10^{-14}$ at most. These search limits will be
powerful tools to probe SUSY at scale of order several TeV.
Therefore, in case of negative measurements for all these LFV
processes, a very sever constraint is expected to be imposed on
the SUSY parameter space.

In minimal supersymmetric seesaw model (which consists of MSSM and right-handed neutrinos),
sizable rates for LFV may be obtained through slepton flavor mixing induced radiatively
by the large neutrino mixing during the evolution from the grand unification (GUT) scale
down to right-handed neutrino scale. In this case, even if universal soft SUSY breaking
parameters are assumed, one finds that the slepton mass matrix receives flavor dependent
radiative corrections and the lepton mass insertionsare given by%
\bea%
(\delta^l_{LL})_{12}  &\sim & \frac{m_0^2}{\tilde{m}^2}
\left(Y_\nu^+ Y_\nu
\right)_{12},~~~~~~~ (\delta^l_{LR})_{12}  \sim  \frac{m_e
A_0}{\tilde{m}^2}\left(Y_\nu^+
Y_\nu \right)_{12}.%
\eea%

For neutrino Yukawa couplings of order one, the above contribution
could enhance the lepton mass insertion significantly. In this
case, the upper bound given in Eq.(\ref{bounds}), in particular
$(\delta^l_{LL})_{12} < 10^{-3}$ is violated unless the slepton
masses are quite heavy. It has been explicitly checked that if the
neutrino Yukawa coupling is of the form: $Y_{\nu} = U_{MNS}
Y^{diag}_{\nu}$, then the predicted SUSY contribution to $\mu \to
e \gamma$ is enhanced significantly and exceeds the current
experimental limits for most of the parameter space
\cite{Calibbi:2006nq}\footnote{See also Ref.\cite{Casas:2001sr}}.
In this respect, the new upper bound $BR(\mu \to e \gamma) \lsim
10^{-13}$ from MEG experiment might impose a lower bound on the
SUSY spectrum of order few TeV, which will be unaccessible at LHC.
Therefore, LFV is a serious test for the large scale seesaw
mechanism within the SUSY framework.

It is therefore of considerable interest to study TeV scale
seesaw, which can easily overcome the LFV problem in SUSY seesaw
models. As shown in the previous section, SUSY
$B-L$ extension of the SM is natural framework for implementing
TeV scale seesaw. In this class of models, the sever constraints
from charged LFV processes are relaxed.

\subsection{$l_i \to l_j  \gamma$ processes}
\begin{figure}[t]
\begin{center}
\epsfig{file=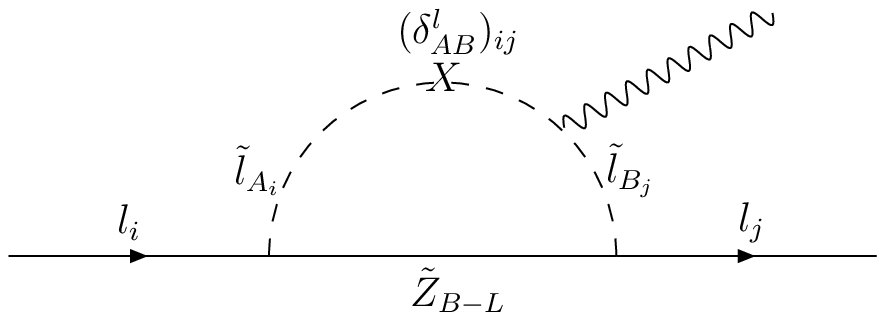, width=7.5cm,height=3.5cm,angle=0}~~~~
\epsfig{file=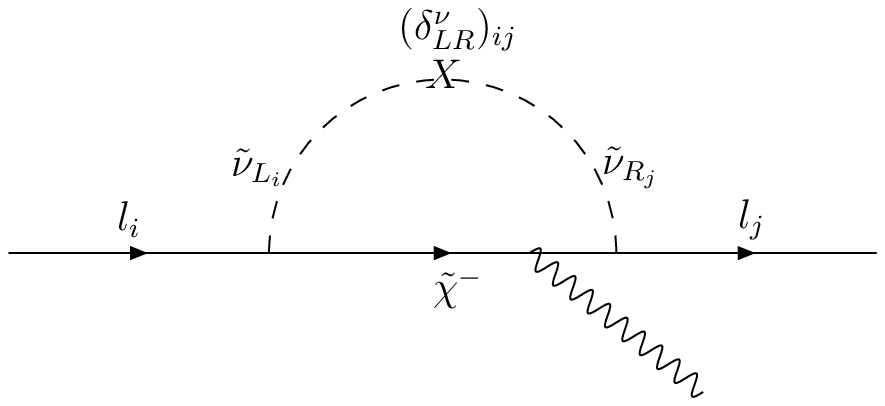, width=7.5cm,height=3.5cm,angle=0}
\end{center}
\vspace{-0.5cm} \caption{New contributions to the decay $l_i \to
l_j \gamma$ in SUSY $B-L$ model.} \label{fig:3}
\end{figure}

In SUSY $B-L$, there are two additional one-loop diagrams
contributing to the decay $l_i \to l_j \gamma$ with $B-L$
neutralino and chargino exchange, as shown in Fig. \ref{fig:3}. In
the $\tilde{Z}_{B-L}$ neutralino contribution, the sleptons are
running in the loop. While the chargino diagram involves both
left- and right- sneutrinos. It is worth noting that these new
contributions are similar to the usual MSSM contribution where
neutralino and slepton or chargino and sneutrino are running in
the loop. Therefore, the model independent bound on the mass
insertions in Eq.(\ref{bounds}) remains valid for
$x=(m_{\tilde{Z}_{B-L}}/m_{\tilde{l}})^2 \simeq 1$. Moreover,
since the soft SUSY breaking terms are now evolving from GUT to
TeV scale, a factor of order ${\cal O}(10)$ is obtained form the
$\ln(M_{GUT}/M_R)$ in the slepton/sneutrino mass corrections.
However, as one can see from Fig. (\ref{fig:3}), these
contributions are proportional to the square of Dirac neutrino
Yukawa. Therefore, they are expected to be quite small.

In fact, the $\tilde{Z}_{B-L}$ neutralino contribution is
proportional to $B-L$ gauge coupling squared time the mass
insertion $(\delta^l_{LL})_{ab}$, which is proportional to
$Y_{\nu}^2$. As emphasized above, in TeV scale seesaw, the Dirac
neutrino Yukawa coupling $Y_\nu$ is of order ${\cal O}(10^{-6})$.
Therefore, the new neutralino amplitude is of order ${\cal
O}(10^{-12})$, which leads to a negligible contribution.

The chargino contribution may be dominated by the mixing between
left- and right-sneutrinos. Note that in large scale SUSY seesaw,
the right-handed (s)neutrinos are decoupled, hence the chargino
contribution is associated with the left-sneutrino only with mass
insertion $(\delta^{\nu}_{LL})_{ij}$ correlated with the
constrained $(\delta^l_{LL})_{ij}$. Thus, in SUSY $B-L$, the new
chargino contribution is given it terms of $SU(2)$ gauge coupling
$g_2$, Dirac neutrino Yukawa coupling $Y_\nu$, and the mass
insertion $(\delta^\nu_{LR})_{12}$. Nevertheless
$(\delta^\nu_{LR})_{12}$ is given by %
\be%
(\delta^\nu_{LR})_{12} \simeq  (Y_{\nu})_{12}~ \frac{v v'}{\tilde{m}^2}. %
\ee%
For $Y_\nu \sim {\cal O}(10^{-6})$, the mass insertion
$(\delta^\nu_{LR})_{ab}$ is of order ${\cal O}(10^{-7})$, hence
the chargino contribution is also quite negligible, of order
${\cal O}(10^{-14})$. Thus, one can conclude that the LFV
associated to $l_i \to l_j \gamma$ processes, which is generated
by RGE from GUT to seesaw scale, is very tiny in SUSY $B-L$ model.

%
\subsection{Same-sign and different flavor dilepton signal at the LHC}
It is important to note that within MSSM or SUSY seesaw
model, another test for LFV at the LHC may be provided by generating
final state with different lepton flavors. For example, $\mu^+$
and $e^- $ can be generated at the final state as follows: $q
\bar{q} \to \tilde{l}_i^+ \tilde{l}_i^- \to \mu^+ e^- + 2
\tilde{\chi}^0$. However, the cross section of this process is
proportional to mass insertion $(\delta^l_{LL})_{12}$. Therefore,
it is typically less than $1$ fb, for slepton mass of order 200
GeV \cite{Deppisch:2004pc}. Note that the dilepton associated with
this process has opposite sign of electric charges. In MSSM or
SUSY seesaw model, same-sign dilepton may be generated only
through the gluino and/or squark production followed by several
cascade decays \cite{Barnett:1993ea}.

Now we consider the same-sign and different flavor dilepton
production mediated by right-handed neutrino and right sneutrino
at the LHC in SUSY $B-L$ model. In particular, we work out the
following two processes, shown in Fig. \ref{fig:4}:
\begin{figure}[t]
\begin{center}
\epsfig{file=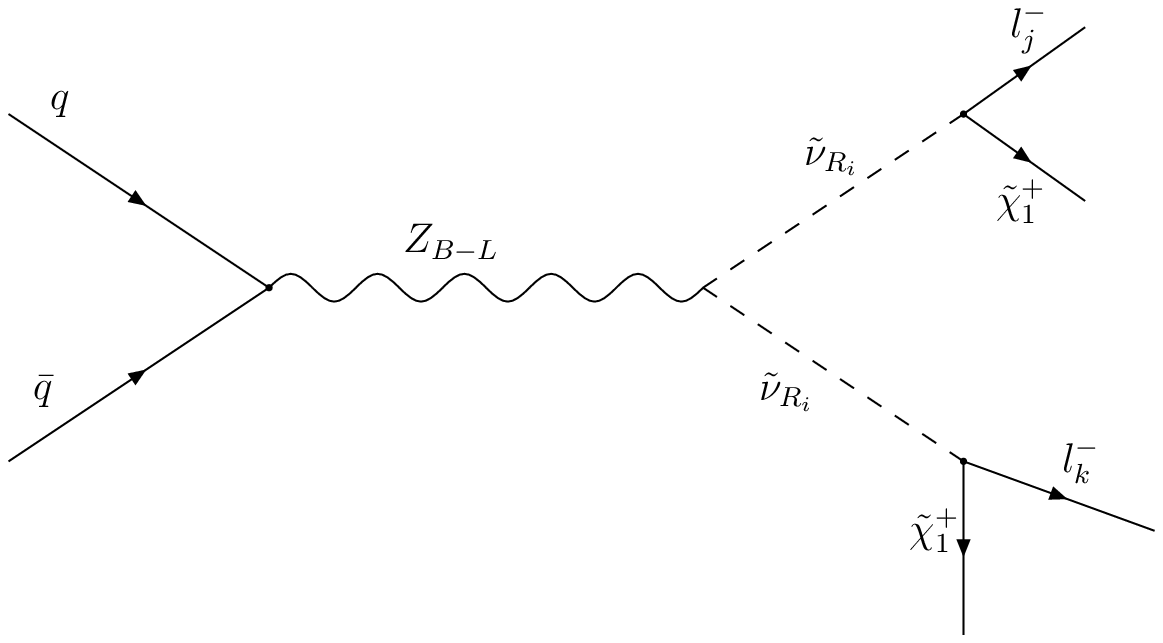, width=6.5cm,height=4.5cm,angle=0}~~~~~~
\epsfig{file=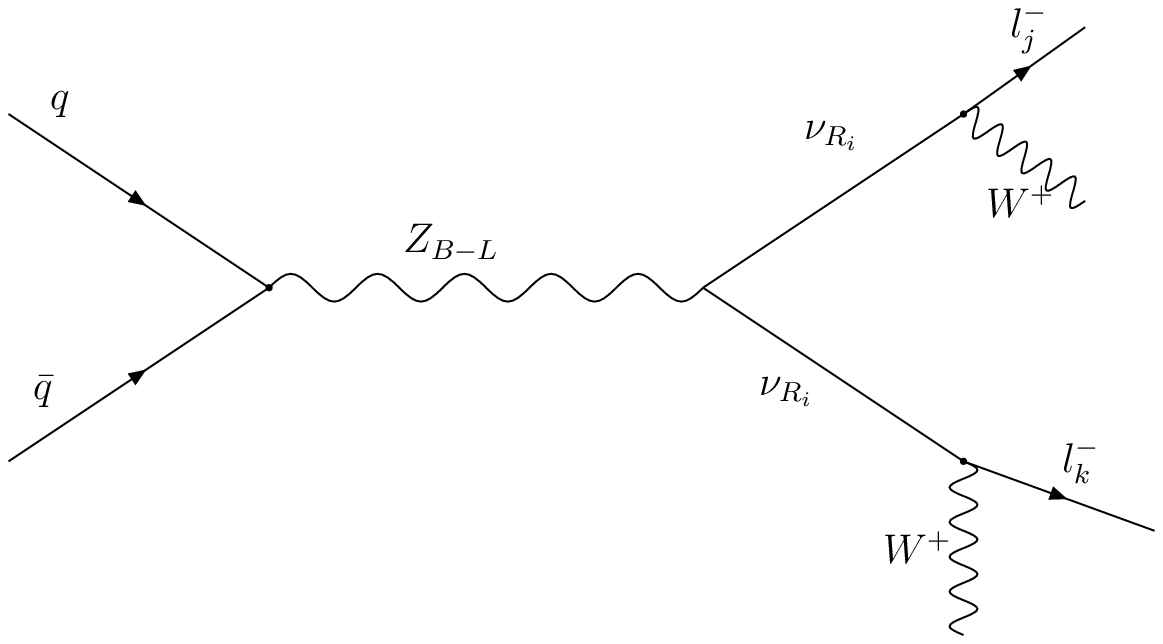, width=6.5cm,height=4.5cm,angle=0}
\end{center}
\vspace{-0.5cm} \caption{Same sign and different lepton flavor
dilepton at the LHC in SUSY $B-L$ model.} \label{fig:4}
\end{figure}
\bea%
&(i)& p p \to \tilde{Z}_{B-L} \to \tilde{\nu}_{R_i}
\tilde{\nu}_{R_i} \to l^-_{j} \chi^+ + l^-_{k} \chi^+ \to l^-_{j}
l^-_{k} + {\rm jets}+ {\rm
missing~ energy}.\nonumber\\
&(ii)& p p \to \tilde{Z}_{B-L} \to \nu_{R_i} \nu_{R_i} \to l^-_{j}
W^+ + l^-_{k} W^+ \to l^-_{j} l^-_{k} + {\rm jets} .\nonumber %
\eea%
Here, the following remarks are in order. $(i)$ In the first
process the LFV is obtained through the right-sneutrino mixing
matrix: $(X_{\tilde{\nu}})_{ij}$. While in the second channel, the
neutrino mixing matrix $U_{ij}$ is the responsible for such flavor
violation. $(ii)$ Both couplings of $\tilde{\nu}_R-l^- -\chi^+$
and $\nu_R-l^- -W^+$ interactions are suppressed by the mixing
between left- and right- neutrino, which is given by: $\sim
m_D/M_R \simeq {\cal O}(10^{-7})$. However, the sneutrino coupling
has another suppression factor $\sim {\cal O}(0.1)$, due to the
chargino diagonlizing matrix, $U_{\chi}$. $(iii)$ One can show
that the decay width of right-sneutrino $\Gamma_{\tilde{\nu}_R}$ and right-handed neutrino $\Gamma_{\nu_R}$
are given by %
\bea%
\Gamma_{\tilde{\nu}_R} \sim  \frac{1}{8\pi}~  \frac{\vert U_{\chi}
Y_{\nu} \vert^2}{m_{\tilde{\nu}_R}}, ~~~~~~~~~~~~ \Gamma_{\nu_R}
\sim \frac{1}{8\pi}~ \frac{\vert Y_{\nu} \vert^2}{m_{\nu_R}}.%
\label{nuwidth}
\eea%
It is clear that $\Gamma_{\tilde{\nu}_R} < \Gamma_{\nu_R}$,
therefore the right-sneutrino is a long-lived particle more than
the right-handed neutrino. In this case, it is expected that the
right-sneutrino will have interesting features at the LHC.
Accordingly, we will focus our discussion on the case of
right-sneutrino pair production.

From the interaction terms in the SUSY $B-L$ Lagrangian, one finds
that the dominant production for the right-sneutrino is through
the exchange of $Z_{B-L}$ and its decay is dominated by chargino
channel, so that $BR(\tilde{\nu}_R \to l^- \chi^+) \simeq 1$. In
general, the amplitude of the process $q \bar{q} \to
\tilde{Z}_{B-L} \to \tilde{\nu}_{R_k} \tilde{\nu}_{R_k} \to
l^-_{i} \chi^+ + l^-_{j} \chi^+$, through the $s$-channel, is
given by \cite{ArkaniHamed:1996au}%
\be%
{\cal M}_{ij} = \sum_i {\cal M}_P \frac{i}{q^2
-\tilde{m}^2_{\tilde{\nu}_{R_k}} + i \tilde{m}_{\tilde{\nu}_{R_k}}
\Gamma_{\tilde{\nu}_{R_k}}} (X_{\tilde{\nu}_R})_{ki}~ {\cal M}_D
\frac{i}{q^2 -\tilde{m}^2_{\tilde{\nu}_{R_k}} + i
\tilde{m}_{\tilde{\nu}_{R_k}}
\Gamma_{\tilde{\nu}_{R_k}}} (X_{\tilde{\nu}_R})^*_{kj}~ {\cal M}_D , ~~~~%
\ee%
where ${\cal M}_P$ is the production amplitude for $q \bar{q} \to
\tilde{\nu}_{R_k} \tilde{\nu}_{R_k}$ and ${\cal M}_D$ is the decay
amplitude for $\tilde{\nu}_{R_k}$. As emphasized in
Ref.\cite{ArkaniHamed:1996au}, the total cross section
$\sigma_{ij}=\sigma(q\bar q\rightarrow \tilde{\nu}_{R_k}
\tilde{\nu}_{R_k} \to l_i^- l_j^- + {\rm jets} + {\rm missing~
energy})$ can
be written as %
\bea %
\sigma_{ij} &=& \int d^2 q \sum_{kl} (X_{\tilde{\nu}_R})_{ki}
(X_{\tilde{\nu}_R})_{kj}^* (X_{\tilde{\nu}_R})_{li}
(X_{\tilde{\nu}_R})_{lj}^* A_{kl}(q^2) \times \left[{\rm
production~
cross~ section}\right]\nonumber\\
&\times& \left[{\rm decay~ branching~ ratio}\right], %
\eea%
where $A_{kl}(q^2)$ is the product of right-sneutrino propagators:
\be%
A_{kl}(q^2) =\frac{i}{q^2 -\tilde{m}^2_{\tilde{\nu_{R_k}}}+ i
\tilde{m}_{\tilde{\nu_{R_k}}} \Gamma_{\tilde{\nu}_{R_k}}}
\frac{i}{q^2 -\tilde{m}^2_{\tilde{\nu}_{R_l}}+ i
\tilde{m}_{\tilde{\nu}_{R_l}}
\Gamma_{\tilde{\nu}_{R_l}}}~.%
\ee%

Assuming that the off-diagonal elements of the right-sneutrino
mass matrix is less than the average right-sneutrino mass, which
is quite natural assumption and always valid in standard SUSY
breaking mechanisms. Also since the decay width
$\Gamma_{\tilde{\nu}_{R}} \ll \tilde{m}_{\tilde{\nu}_R}$, one can
approximate the
right-sneutrino propagator as follows \cite{Kalinowski:2001tq}:%
\be%
A_{kl}(q^2) =\frac{1}{1 + i \Delta
M_{\tilde{\nu}_R}/\Gamma_{\tilde{\nu}_{R_k}}} ~
\frac{\pi}{m_{\tilde{\nu}_{R}}\Gamma_{\tilde{\nu}_{R}}}~ \delta\left(q^2 -m_{\tilde{\nu}_R}^2\right)~.%
\ee%
Thus, the total cross section $\sigma_{ij}$ can be written as %
\cite{Kalinowski:2001tq}%
\be%
\sigma_{ij}\approx \frac{\vert (\Delta
m^2_{\tilde{\nu}_R})_{ij}\vert^2}{\tilde{m}_{\tilde{\nu}_R}^2
\Gamma_{\tilde{\nu}_R}^2}~ \sigma(q\bar q\rightarrow \tilde{\nu}_R
\tilde{\nu}_R), %
\ee%
From Eq.(\ref{nuwidth}), the right-sneutrino decay width is of
order $\Gamma_{\tilde{\nu}_R} \lsim {\cal O}(10^{-14})~ {\rm
GeV}^{-1}$. Therefore, with $m_{\tilde{\nu}_R} \sim {\cal O}(1)~
{\rm TeV}$ and $\Delta M_{\tilde{\nu}}/m_{\tilde{\nu}_R} \sim
{\cal O}(10^{-2})$, one finds \footnote{More detailed analysis for
these processes at LHC, with specific models of supersymmetry
breaking,
will be considered elsewhere}%
\be%
\sigma_{ij}\approx 10^{10} ~ \sigma(q\bar q\rightarrow \tilde{\nu}_R\tilde{\nu}_R), %
\ee %
with
\be %
\sigma(q\bar q\rightarrow
\tilde{\nu}_R\tilde{\nu}_R)\simeq\frac{g^4_{B-L}m^2_q}{6\pi
\left(s^2-m^2_{Z_{B-L}}\right)^2} \sqrt{ 1-
\left(\frac{2m_{\tilde{\nu} _R}}{s}\right)^2}\left[ 1-
\left(\frac{2m_q}{s}\right)^2  \right]. %
\ee%
It is remarkable that for gauge coupling $g^{\prime\prime} \sim
{\cal O}(0.1)$ and $m_{Z_{B-L}} \sim {\cal O}(1)$ TeV one finds
that the
cross section is given by %
\be%
\sigma(q\bar{q} \to l_i^- l_j^- + {\rm jet}+ {\rm missing ~energy}) \simeq 10^{-10}~ {\rm GeV}^{-2}
\simeq {\cal O}(1)~ {\rm pb}~ .%
\ee%
For these values of cross section, the same-sign dilepton signal
can be easily probed at the LHC. This event will be a clear hint
for sizeable LFV at the LHC, which is more significant than the
bounds obtained from the rare decays, $l_i \to l_j \gamma$. Since
the SM background of the same-sign dilepton is negligibly small,
the discovery of this process would be undoubted signal for SUSY
$B-L$ model.

\section{Conclusions}

We have analyzed the LFV in supersymmetric $B-L$ extension of the
standard model. In this model, $B-L$ symmetry is radiatively
broken at TeV scale. Therefore, it is a natural framework for TeV
scale seesaw mechanism with Dirac neutrino Yukawa coupling of
order ${\cal O}(10^{-6})$. We have shown that because of the
smallness of Dirac neutrino Yukawa couplings, the decay rates of
$l_i \to l_j \gamma$ and $l_i \to 3 l_j$, generated by the RGE
from GUT to seesaw scale, are quite suppressed. In this case, the
LFV constraints imposed on this class of models remain as in the
MSSM. Also, we have studied another possibility for LFV at the
LHC, which associated with the same-sign dilepton, produced
through the decay of the long lived pair of right-sneutrinos. We
have shown that the total cross section of the process: $q \bar{q}
\to Z_{B-L} \to \tilde{\nu}_{R_i} \tilde{\nu}_{R_i} \to l_j^{\pm}
l_k^{\pm} + {\rm jets} + {\rm missing~ energy}$ is of order ${\cal
O}(1)$ pb. Therefore, it is experimentally accessible at the LHC,
with negligibly small SM background. The probe of this signal will
provide indisputable evidence for SUSY $B-L$ extension of the SM
and also for right sneutrino-anti-sneutrino oscillation.

\section{Acknowledgments}
I would like to thank H. Okada for fruitful discussions. I am very
grateful to the members of the NExT institute at Southampton
University and STFC, Rutherford Appleton Laboratory for their kind
hospitality, where this work was completed. We also acknowledge
financial support through the Centre for Fundamental Physics. This
work was partially supported by the Science and Technology
Development Fund (STDF) Project ID 437 and the ICTP Project ID 30.


\end{document}